\documentstyle[12pt]{article}

\renewcommand{\baselinestretch}{1.3}

\def\({\c c}
\def\|{\'\i}

\newcommand{\VEV}[1]{\left\langle{#1}\right\rangle}

\begin{document}

\baselineskip=15pt   % note it

\begin{flushright}
SLAC-PUB-8168\\
MAY 1999\\
%hep-th/
\end{flushright}

\medskip

\baselineskip=18pt  %

\begin{center}
{\large\bf LIGHT-FRONT-QUANTIZED QCD IN COVARIANT GAUGE}
\footnote{Research
partially supported by the Department of Energy under
contract DE-AC03-76SF00515 }

\vspace{1cm}

\baselineskip=20pt  %
{\large Prem P. Srivastava}\footnote{E-mail:\quad prem@slac.stanford.edu
or prem@cbpfsu1.cat.cbpf.br. On
leave of absence from  \\
{\it Instituto de F\'{\i}sica, UERJ-Universidade do
Estado de Rio de Janeiro, Brasil}.} \\
and \\
{\large Stanley J. Brodsky}\footnote{E-mail:\quad sjbth@slac.stanford.edu}

\vspace{0.3cm}
{\em \it Stanford Linear Accelerator Center, Stanford University,
Stanford, California 94309}
\vspace{0.2cm}

{\bf Abstract}

\end{center}

\baselineskip=17pt
{\small \quad The light-front (LF) canonical quantization of quantum
chromodynamics in covariant gauge is discussed. The Dirac procedure is 
used to eliminate the constraints in the gauge-fixed {\it front form} 
theory quantum action and to construct the LF Hamiltonian formulation. 
The physical degrees of freedom emerge naturally. 
 The propagator of the dynamical 
 $\psi_+$ part of the free fermionic
propagator in the LF quantized field theory is shown to be causal and not
to contain instantaneous terms.
   Since the relevant 
 propagators in the covariant gauge
formulation are causal,  
 rotational invariance---including  the Coulomb potential in the static 
 limit---can be recovered, avoiding the difficulties 
encountered in light-cone gauge.  The Wick rotation may also be performed 
allowing the conversion of momentum space integrals into Euclidean
space forms.   Some
explicit computations are done in quantum electrodynamics to
illustrate the equivalence of {\it front form} theory with the
conventional covariant formulation.  LF quantization thus provides a
consistent formulation of gauge theory, despite the fact that the
hyperplanes $x^{\pm}=0$ used to impose boundary conditions 
constitute characteristic
surfaces of a hyperbolic partial differential equation.}
 \vspace{0.2cm}

\begin{center}
(Submitted to Physical Review D.)
\end{center}

%\nl {\bf Keywords}:\quad{Light-front quantization, Gauge Theory, Covariant
%gauges, Propagator}

\newpage

\renewcommand{\baselinestretch}{2}

\section{Introduction}\label{intrO}

The quantization of relativistic field theory at fixed light-front time
$\tau = (t - z/c)/\sqrt 2$,  which was proposed by Dirac \cite{dir} half a
century ago, has found important applications \cite{bro,ken,pre} in
gauge theory and string theory \cite{{susskind}}.
   The
light-front (LF) quantization of QCD in its Hamiltonian form provides an
alternative approach to lattice gauge theory for the computation of
nonperturbative quantities, such as the
spectrum and the light-cone Fock state wavefunctions of relativistic bound
states.  LF variables have found natural applications in several context, 
for example, in 
the quantization of (super-) string theory and M-theory \cite{susskind}.  
It  has also been
 employed in the nonabelian
bosonization \cite{wit} of the field theory of $N$ free Majorana fermions.  
 
Since LF coordinates are not related to the conventional
coordinates  by a finite Lorentz
transformation, the descriptions of the same physical
result may be different in the equal-time ({\it instant form}) 
and equal LF-time  ({\it front form}) 
of treatments of  the theory.  This was also  found to be the case  in the
recent study \cite{pre1}  of some soluble two-dimensional gauge theory models,
where it was also demonstrated 
that LF quantization is very economical in displaying the
relevant degrees of freedom, leading  directly to a physical
Hilbert space.      LF-time-ordered
perturbation theory has  also been applied \cite {soper, yan}  
to  massive fields.   It was used  in the analysis of the 
evolution of deep inelastic
structure functions \cite{Bjorken:1971ah},   the evolution of
the distribution amplitudes which control hard exclusive processes in
       structure functions \cite{Bjorken:1971ah}, and the evolution of
the distribution amplitudes which control hard exclusive processes in
QCD \cite{Lepage:1980fj}. LF-time-ordered perturbation theory is
much more economical than  equal-time-ordered perturbation
theory, since only graphs with particles with positive LF momenta
$p^+ = (p^0 + p^3)/\sqrt 2$ appear.
 It has been conventional to apply light-front quantization to gauge
theory in light cone gauge $A^+ = (A^0 + A^3)/\sqrt 2=0$, since the
transverse degrees of freedom of the gauge field can  be immediately
identified  as the dynamical degrees of freedom, and the ghost fields  
can be ignored  in the quantum action of the nonabelian 
gauge theory \cite{Bassetto:1997ba}.  
  However, it does 
 bring severe penalties, such as the breaking of manifest rotational
invariance, cumbersome LF time instantaneous interactions, an
analytically complex gauge-field propagator, and a difficult
renormalization procedure \cite{{Perry:1999sh}}.  In addition, 
light-cone gauge complicates
obtaining the Coulomb limit when one applies the LF formalism to simple
nonrelativistic systems.
 
In this paper we will discuss the LF quantization of (massless) 
 gauge field theory in
 covariant gauges such as Feynman gauge. 
 The penalty in reintroducing Faddeev-Popov ghosts  and other  fields with
ghost-like metric is compensated by the restoration 
of the Lorentz symmetries.  Our general procedure will be first to 
study the  gauge-fixed quantum action of the theory on the LF. The 
self-consistent Hamiltonian 
formulation is  constructed following the Dirac method 
\cite{dir1} for constrained systems. 
It is worth remarking that even after   the addition of  
gauge-fixing and ghost terms   in the {\it front form} 
theory, several   second class constraints in general  remain in the theory.  
They can be   eliminated  by constructing the Dirac brackets, and the   
theory  can be quantized canonically by the correspondence principle in terms of a
{\it reduced number} of independent fields. The commutation relations 
among them are also found by the Dirac method, and they  are useful in order to 
obtain the momentum space expansions of the fields. The nondynamical
projection of the fermion field will be eliminated using a nonlocal
constraint equation.   The 
 Dyson-Wick perturbation theory expansion \cite{bjk} based on 
LF coordinates then becomes   straightforward \cite{yan}. 
In fact, the Feynman gauge has been already  used \cite{bro2} in QED in
the framework of Weinberg's time-ordered perturbation theory in the
infinite-momentum-frame, where it was used to renormalize and compute the
order $\alpha, \alpha^2$ and part of the $\alpha^3$ radiative
corrections to the electron anomalous moment. (In this gauge the theory
on the LF resembles that of an interacting theory of four {\sl scalar} fields 
$(A^+, A^\perp, A^-)$ with the matter fields. 
The covariant gauge thus  should be feasible within the LF framework.) 
The challenge of 
 renormalization in light-cone gauge and some of its aspects in 
 Feynman gauge were addressed in \cite{burkardt}.      
   The {\it front form} theory 
has recently been used \cite{Miller:1998tp}  to  
quantize a nuclear
physics model in which massive vector mesons 
couple to field-theoretic
nucleons.  In this approach, a canonical transformation \cite {soper, yan} 
of the dynamical nucleon
$\psi_+$ field is used to simplify the constraint equation for $\psi_-$ 
(in abelian theory). 
The inclusion of fields with ghost-like metric in discretized
light-cone quantization (DLCQ) \cite{Pauli:1985ps} has been discussed in
\cite{Brodsky:1999xj}.

The propagator for the dynamical component  $\psi_+(x)$ 
of the free fermionic field
 on the LF is shown to be  causal,  and it is demonstrated that, 
 in the context of LF-time-ordered Dyson-Wick perturbation theory,  it 
  has no instantaneous term. 
   This effect \cite{Lepage:1980fj} is 
replaced now by explicit seagull and other interaction terms  which can be
evaluated systematically.   Such interactions can be
incorporated into a nonperturbative approach such as DLCQ. 
  Next QCD
is canonically quantized on the LF in  covariant Feynman (or Landau)
gauge. A novel feature of the gauge theory interactions in covariant
gauges on the LF is  the off-diagonal couplings of the
$A^+$ and
$A^-$ fields. Some illustrations are  given  on the  
 application the Dyson-Wick perturbation theory expansion 
based on the Wick  theorem, built on the LF-time-ordered product.     
We recall that it was also used \cite{pre2} to 
renormalize two-dimensional scalar field theory  on the LF, with 
nonlocal interaction.  It was
also  shown  there \cite{pre2} that even on the
LF one may perform the  Wick rotation in momentum
space integrals and go over to the
Euclidean-space integrals, rendering the
computation of  high order
corrections as straightforward as in conventional theory.

\section{Notation}

The Lagrangian density corresponding to the quantum action \cite{nak} 
 of  QCD is described in  standard notation  by 
\begin{equation}
{\cal L}_{QCD}=-{1\over 4}F^{a\mu\nu}{F^{a}}_{\mu\nu}+B^{a}\partial_{\mu}A^{a\mu}+
{{\xi} \over 2}B^{a}B^{a}+i\partial^{\mu}{\chi_{1}}^{a}{{\cal D}^{ac}}_{\mu}
{\chi_{2}}^{c}+ {\bar\psi}^{i}
(i\gamma^{\mu}{D^{ij}}_{\mu}-m\delta^{ij})\psi^{j}
\end{equation}
Here $\psi^{j}$ is the quark field with color index $j=1..N_{c}$ for an  
$SU(N_{c})$ color group, ${A^{a}}_{\mu}$ the gluon field,   
 $F^{a}_{\mu\nu}=\partial_{\mu}A^{a}_{\nu}
-\partial_{\nu}{A^{a}}_{\mu}
+g f^{abc}{A^{b}}_{\mu}{A^{c}}_{\nu}$ the  
field strength, ${{\cal D}^{ac}}_{\mu}=
(\delta^{ac}\partial_{\mu}+g f^{abc}{A^{b}}_{\mu})$,  
${D^{ij}}_{\mu}\psi^{j}=
(\delta^{ij}\partial_{\mu}-ig {A^{a}}_{\mu}(\lambda^{a}/2)^{ij})\psi^{j} $, 
$a=1..({N_{c}}^{2}-1)$  the gauge group 
index etc. The covariant  
gauge-fixing is introduced  by adding to the Lagrangian 
the linear gauge-fixing term $(B^{a}\partial_{\mu}A^{a\mu}+
{{\xi} \over 2}B^{a}B^{a})$ where $B^{a}$ is 
the  Nakanishi-Lautrup  auxiliary field 
and $\xi$ is a parameter.  
The ${\chi_{1}}^{a}$ and 
${ \chi_{2}}^{a}$ are the (hermitian) anticommuting Faddeev-Popov 
ghost fields, and the   action is 
invariant under the BRS \cite{stora} transformation. 

The LF   
coordinates  $\,\mu=+,-,1,2\,$ are defined as 
$x^{\mu}=(x^{+}=x_{-}=(x^{0}+x^{3})/{\sqrt 2},\, x^{-}=x_{+}=
(x^{0}-x^{3})/{\sqrt 2},\, x^{\perp})$ where $x^{\perp}= (x^{1},x^{2})
=(-x_{1},-x_{2})$ 
are the transverse coordinates and $\,\mu=+,-,1,2$.   
The  $x^{+}\equiv \tau$ will be taken as 
the LF-time coordinate while $x^{-}$ is the longitudinal spatial one. We 
can of course make the convention with the role of $x^{+}$ and $x^{-}$ 
interchanged. The equal-$x^{+}$ quantized theory does seem to already contain
the information on the equal-$x^{-}$ commutator \cite{pre1}.  For the  
 four-momentum components we have  $k^{\mu}=
 (k^{-}\equiv(k^{0}-k^{3})/{\sqrt 2}, \,k^{+}\equiv
(k^{0}+k^{3})/
{\sqrt 2},\, k^{\perp})$ where $k^{+}$ indicates the  
longitudinal momentum  and  $k^{-}$ 
 the LF energy. The LF components of any tensor, for example,  the gauge field 
$(A^{+}, A^{-}, A^{\perp})$ are similarly defined, and the  metric tensor 
$g_{{\mu}{\nu}}$ may be read from  
 $A\cdot B= A^{+}B^{-}+A^{-}B^{+}
-A^{\perp}B^{\perp}$. The $\gamma^{\pm}$  defined by  
   $\gamma^{\pm}=(\gamma^{0}\pm \gamma^{3})/{\sqrt {2}}$ satisfy  
$({\gamma^{+}})^{2}=({\gamma^{-}})^{2}=0$, 
$\gamma^{0}\gamma^{+}=\gamma^{-}\gamma^{0}$, 
$\gamma^{+}\gamma^{-}+ 
\gamma^{-}\gamma^{+}=2 I$, ${\gamma^{+}}^{\dag}={\gamma^{-}}$ and  
 it follows that 
$\,\Lambda^{\pm}= {1\over 2} \gamma^{\mp}\gamma^{\pm}= {1\over \sqrt {2}}
\gamma^{0}\gamma^{\pm}\,$ are hermitian projection operators: $\, 
\quad({\Lambda^{\pm}})^{2}={\Lambda^{\pm}}, \quad {\Lambda^{+}}
\Lambda^{-}=\Lambda^{-}\Lambda^{+}=0,
\quad \gamma^{0}\Lambda^{+}=\Lambda^{-}\gamma^{0}$.   
The corresponding $\,{\pm}\,$ projections of the LF Dirac spinor 
are $\psi_{\pm}=\Lambda^{\pm}\psi$ and  $\psi=\psi_{+}+\psi_{-},  
\bar\psi = \psi^{\dag}\gamma^{0}= \bar\psi_{+}+\bar\psi_{-}$, 
$\gamma^{\pm}\psi_{\mp}=0$ etc. 

\section{Spinor field propagator on the LF}

The quark field term in  LF coordinates reads 
\begin{eqnarray}
{\bar\psi}^{i}
(i\gamma^{\mu}{D^{ij}}_{\mu}-m\delta^{ij})\psi^{j}
&=& i{ \sqrt {2}}{\bar\psi_{+}}^{i}
\gamma^{0}{D^{ij}}_{+}{\psi_{+}}^{j}+{\bar\psi_{+}}^{i}
(i\gamma^{\perp}{D^{ij}}_{\perp}-m\delta^{ij}){\psi_{-}}^{j} \nonumber \\
&+&{\bar\psi_{-}}^{i}\left[{i \sqrt {2}}\gamma^{0}
{D^{ij}}_{-}{\psi_{-}}^{j}+ (i\gamma^{\perp}{D^{ij}}_{\perp}-m\delta^{ij})
{\psi_{+}}^{j}\right]
\end{eqnarray}
where ${D^{ij}}_{\pm}=
(\delta^{ij}\partial_{\pm}-ig {A^{a}}_{\pm}(\lambda^{a}/2)^{ij}) $.                              
It shows that the  minus components ${\psi_{-}}^{j}$  are in fact nondynamical 
( Lagrange multiplier ) fields 
without  kinetic terms. 
The variation of the action with respect to  
${{\bar{\psi^{j}}}_{-}}$ and ${{{\psi^{j}}}_{-}}$ leads to 
the following gauge covariant constraint equation 
\begin{equation}
i{\sqrt 2} {D^{ij}}_{-}{\psi_{-}}^{j}= -(i\gamma^{0}\gamma^{\perp}{D^{ij}}_{\perp}-m\gamma^{0}\delta^{ij})
{\psi_{+}}^{j}, 
\end{equation}
and its conjugate.  
The   ${\psi^{j}}_{-}$ components may thus be eliminated in favor of 
 the dynamical components $\psi_{+}^{j}$
\begin{equation}
{\psi_{-}}^{j}(x)
= \frac{i}{\sqrt 2}  \left[U^{-1}(x|A_{-})
\frac{1}{\partial_{-}}
U(x|A_{-})\right]^{jk}
(i\gamma^{0}\gamma^{\perp}{D^{kl}}_{\perp}-m\gamma^{0}\delta^{kl})
{\psi_{+}}^{l}(x). 
\end{equation}
Here, for a fixed $\tau$,  $U\equiv U(x|A_{-})$ is an $N_{c}\times N_{c}$ 
gauge matrix  in the fundamental representation of $SU(N_{c})$ and 
it satisfies   
\begin{equation}
\partial_{-}U(x|A_{-})= -ig \,U(x|A_{-})\, A_{-}(x)
\end{equation}
with  $A_{-}={A^{a}}_{-}\lambda^{a}/2$. It has  
 the formal solution 
\begin{equation}
U(x^{-},x^{\perp}|A_{-})=U({x^{-}}_{0},x^{\perp}|A_{-})\,\tilde {\cal P} 
\,exp\left\{-ig \int_{{x^{-}}_{0}}^{x^{-}} dy^{-}
 A_{-}(y^{-},x^{\perp})\right\}
\end{equation}
where $\tilde {\cal P}$ indicates the anti-path-ordering along the 
longitudinal direction $x^{-}$.  $U$ has  a series expansion 
in the powers of the coupling constant.

The free  field 
propagator for $\psi_{+}$  is determined from 
the following quadratic terms in the  Lagrangian (suppressing the 
color index) 
density 
\begin{equation}
 i\sqrt {2} {\psi_{+}}^{\dag}\partial_{+}\psi_{+}
+{\psi_{+}}^{\dag}(i\gamma^{0}\gamma^{\perp}\partial_{\perp}- 
m\gamma^{0})\psi_{-}
\end{equation}
contained in  (2). Here we  have  used the free field constraint equation
$\,2i\partial_{-}\psi_{-}=(i\gamma^{\perp}\partial_{\perp}+m)
\gamma^{+}\psi_{+}\,$ which determines the 
dependent field $\psi_{-}$.   The equation of motion for the independent 
component $\psi_{+}$ is nonlocal in the longitudinal direction
\begin{equation}
\left[4\partial_{+}+ (m+ i\gamma^{\perp}\partial_{\perp} )
 \gamma^{-}{1\over{\partial_{-}}}
(m+ i \gamma^{\perp}\partial_{\perp}) \gamma^{+} \right] \psi_{+}=0. 
\end{equation}
The free field   Hamiltonian formulation  can be constructed   by following 
the  Dirac procedure \cite{dir1}.       
The   constraint equation   arises now as 
 a  second class constraint on the canonical phase space. 
  The  Dirac bracket which takes care of these constraints is easily 
  constructed.   
The effective free LF Hamiltonian is found to be    $\;{\cal H}^{lf}= 
 -{{\bar\psi}_{+}}(i\gamma^{\perp}\partial_{\perp}-m)\psi_{-}$  
and  the  canonical quantization  performed by the  correspondence 
of the  Dirac brackets with  the (anti-)commutators  
leads to
 the following nonvanishing {\it local} anticommutation relation 
 \begin{equation}
 \{{\psi}_{+}(\tau, x^{-},x^{\perp}),{{\psi}_{+}}^{\dag}(\tau, y^{-},y^{\perp})\}=
 {1\over \sqrt {2}} 
 \Lambda^{+} \delta(x^{-}-y^{-})\delta^{2}(x^{\perp}-y^{\perp}). 
 \end{equation}
on the LF.  They were proposed earlier in Ref. [8].  The   equation of motion (8) for    
$\psi_{+}$ is   recovered   as an Heisenberg equation of motion if we 
employ  (9).

  The  propagator in momentum space may be derived  by going over to  
 the Fourier transform of $\psi(x)$   
 over the complete set of linearly independent plane wave solutions of the free 
 Dirac equation, say, for  $p^{+}>0 $.   
      Such a set is spanned by 
 $ u^{(r)}(p) e^{-i p\cdot x}$ together with 
    $ v^{(r)}(p) e^{i p\cdot x}$ 
where $u^{(r)}(p)$ and  
$v^{(r)}(p)\equiv {u^{(r)}(p)_{c}}= C {\gamma^{0}}^{T} u^{(r)}(p)^{*}$    
are linearly independent solutions of the Dirac equation in momentum space:  
  $(\gamma^{\mu}p_{\mu}-m)u^{(r)}(p)=0$ and $p\cdot x=(p^{-}x^{+}+p^{+}x^{-}-
  p^{\perp}x^{\perp})$.   

A   useful  form \cite{ref:aa} of the solution for  the four-spinors in the context of  
the LF quantization is 
 \begin{equation}
u^{(r)}(p)
= \frac{1}{({ {\sqrt {2}}p^{+} m })^{1\over 2}}\left[{\sqrt 2} p^{+}\Lambda^{+} 
+(m+\gamma^{\perp}p_{\perp})
\Lambda^{-}\right] \tilde u^{(r)} \nonumber \\
\end{equation}
where the  constant spinors $\tilde u^{(r)}$ satisfy 
$\gamma^{0}\tilde u^{(r)}=\tilde u^{(r)}$ and   $ \Sigma_{3}
\tilde u^{(r)}= r \tilde u^{(r)}$ with $\Sigma_{3}= i\gamma^{1}\gamma^{2}$ 
and  $r=\pm $.  The normalization and the completeness relations are: 
\quad $\bar u^{(r)}(p)u^{(s)}(p)=\delta_{rs}=-\bar v^{(r)}(p)v^{(s)}(p)$, 
$\;\sum_{r=\pm} u^{(r)}(p){\bar u}^{(r)}(p)= {(\not p+m)/ {2m}}$,  
$\;\sum_{r=\pm} v^{(r)}(p){\bar v}^{(r)}(p)= {(\not p-m)/ {2m}}$ and 
C is the charge conjugation matrix \cite{bjk}. 
 
Its  $\Lambda^{+} $ projection is by construction very simple,  
${u^{(r)}}_{+}(p)={({\sqrt {2}}p^{+}/m)}^{1\over 2}
(\Lambda^{+}{\tilde u^{(r)}})$  and 
 they  are eigenstates of $\Sigma_{3}$ as well while the 
 $\tilde u^{(r)}$ correspond
to the rest frame spinors for which ${\sqrt 2}p^{\pm}=m$. 

The Fourier transform expansion may be written as
\begin {equation}
\psi(x)={1\over {\sqrt {(2\pi)^{3}}}}
              \sum_{r={\pm}}\int d^{2}p^{\perp}
	      dp^{+}\,
	      \theta(p^{+}){\sqrt{m\over p^{+}}}\left[b^{(r)}(p){ u^{(r)}}(p)
	       e^{-ip.x}+
          d^{{\dag}{(r)}}(p){ v^{(r)}}(p) e^{ip.x}\right]
\end {equation}   
The presence  of the factor 
$\theta(p^{+})$     in (11) follows  also   from the considerations of the 
{\it covariant phase space} (or {\it LIPS}) {\it factor} which is found 
 relevant \cite{ref:bb} in  the context of the 
 analysis  of the physical processes.  
For $\psi_{+}\equiv \Lambda^{+}\psi$ we  find
\begin{equation}
{\psi_{+}}(x)={({\sqrt 2})^{1\over 2}\over {\sqrt {(2\pi)^{3}}}}
              \sum_{r={\pm}}\int d^{2}p^{\perp}
	      dp^{+}
	      \theta(p^{+})\left[b^{(r)}(p){\tilde u_{+}}^{(r)} e^{-ip.x}+
          {d^{\dag}}^{(r)}(p){{\tilde v}_{+}}^{(r)} e^{ip.x} \right], 
\end{equation}	  
where $\tilde u$ and $\tilde v$ are constant   
spinors, and   the integrations are to be taken    
 from $-\infty$ to $\infty$ not only
for $p^{\perp}$ but also over  $p^{+}$. 
It can be  verified that the anticommutation relations (9) 
 are satisfied if the creation and the annihilation operators 
are assumed
to satisfy the canonical anticommutation relations, with the
nonvanishing ones given by: $\;\{b^{(r)}(p),{b^{\dag}}^{(s)}(p')\}
=\{d^{(r)}(p),{d^{\dag}}^{(s)}(p')\}= \delta_{rs}
\delta(p^{+}-p'^{+})\delta^{2}(p^{\perp}-p'^{\perp})$.

The free propagator for the independent component $\psi_{+}$ is 
easily derived  using (12) and (10) and it is given  
by
\begin{eqnarray}
\VEV{0|\,T(\psi_{+ A}(x)\psi^\dag_{+ B}(0))\,|0} 
&=& 
\VEV{0|\left[\theta(\tau){\psi_{+ A}(x)}\psi^\dag_{+ B}(0)
-\theta(-\tau)\psi^{\dag}_{+ B}(0)\psi_{+ A}(x)\right]|0}  
\nonumber \\[1ex]
&=& 
\frac{1}{\sqrt 2}\frac{\Lambda^+_{AB}}{(2\pi)^3}\int 
d^2q^\perp dq^+\theta(q^+)\left[\theta(\tau) e^{-iqx}- 
\theta(-\tau) e^{iqx}\right]
\end{eqnarray}
where $A,B=1.4\,$ label the spinor components and we have used (12). 
The only relevant 
differences, compared with the case of the scalar field, are,  
 apart from the appearance  of the projection operator,  
  the absence of  the factor  
 $(1/2q^{+})$ in the integrand of (13) and the 
 negative sign of  the second term in  the fermionic case. 
 They, however,   compensate, and the standard manipulations to factor out 
 the exponential give rise to the factor 
 $\left[\theta(q^+)+\theta(-q^+)\right]$  which may be interpreted as  
 unity in 
 the distribution theory sense,   parallel to what we find 
 in the derivation of the   scalar field propagator on the LF.    
The straightforward use  of 
 the integral representation $2\pi i \,\theta(\tau)\,e^{-ip\tau}=
 \int d\lambda \,e^{(-i\lambda \tau)}/(p-\lambda-i\epsilon)$ 
  of $\,\theta(\pm \tau)$, together with  
  the standard manipulations in the second term to factor out the exponential,  
    leads to\begin{equation}
<0|T({\psi^{i}}_{+}(x){\psi^{\dag j}}_{+}(0))|0>  =   
{{i\delta^{ij}}\over {(2\pi)^{4}}} \int d^{4}q \;{{{\sqrt {2}}q^{+}\, 
\Lambda^{+} }\over 
{(q^2-m^{2}+i\epsilon)}}\, e^{-iq.x}, 
\end{equation}
where for convenience we have renamed the dummy integration variable 
originating from  the integral representation of 
$\theta(\pm\tau)$ as $q^{-}$   and 
 $d^{4}q=d^{2}q^{\perp}d q^{+} dq^{-}$ with  all  integrations 
 ranging from  $-\infty$ to  $\infty$. It may 
  also be derived straightforwardly by functional integral; we do have to 
 take care of the second class constraint in the measure. 
We have restored the color index 
 as well. The propagator (14) contains   no   instantaneous  
 term \cite{soper},  and the integrand factor  may also be expressed as 
$\approx\,\left [\Lambda^{+} (\not q +m)\Lambda^{-}/
{(q^2-m^{2}+i\epsilon)}\right]\gamma^{0} $. 
We verify   that the propagator   satisfies  the equation for the 
Green's function corresponding to the equation of motion 
of $\psi_{+}$, (8). 

\section{Gauge field propagator}

We now turn our attention to the  gauge field  propagator on the LF  in 
the covariant gauges. The relevant  quadratic terms in the 
Lagrangian density  are  
\begin{equation}
\frac{1}{2}\left[{F^{a}}_{+-}{F^{a}}_{+-}+2{F^{a}}_{+{\perp}}{F^{a}}_{-{\perp}}
 -{F^{a}}_{12}{F^{a}}_{12} \right] 
++B^{a}\partial_{\mu}A^{a\mu}+
{{\xi} \over 2}B^{a}B^{a}+i\partial^{\mu}{\chi_{1}}^{a} {\partial }_{\mu}
{\chi_{2}}^{a}
\end{equation}
It will be  sufficient   to study   the free abelian gauge theory described by  the 
following Lagrangian density 
\begin{equation}
{1\over 2}\left[(F_{+-})^{2}-(F_{12})^2 +2F_{+\perp}
F_{-\perp}\right]+B(\partial_{+}A_{-}+\partial_{-}A_{+}+\partial_{\perp}A^{\perp})
+{\xi\over2}B^{2},
\end{equation}
where for the abelian theory   
$F_{\mu\nu}\equiv (\partial_{\mu}A_{\nu}-\partial_{\nu}A_{\mu})$.  
The canonical momenta 
are  $\pi^{+}=0$, $\pi_{B}=0$, 
$\pi^{\perp}=F_{-\perp}$, $\pi^{-}=F_{+-}+B$ and the canonical 
 Hamiltonian density is found to be 
\begin{equation}
{\cal H}_{c}= {1\over 2} ({\pi}^{-})^{2}+{1\over2}(F_{12})^{2}-
A_{+}(\partial_{-}\pi^{-}+\partial_{\perp}\pi^{\perp}-2\partial_{-}B)
-B(\pi^{-}+\partial_{\perp}A^{\perp})+{1\over2}(1-\xi)B^2
\end{equation}
In the Dirac procedure the primary constraints \cite{dir1} are $\pi^{+}\approx 0$, 
$\,\pi_{B}\approx 0$ and $\,\eta\equiv\pi^{\perp}-\partial_{-}A_{\perp}+
\partial_{\perp}A_{-}\approx 0$, where $\perp=1,2$ and $\,\approx\,$ 
stands for {\it weak equality} relation \cite{dir1}. We now require 
the persistency in $\tau$ of these constraints employing  the 
preliminary Hamiltonian, which is obtained by adding to the canonical 
Hamiltonian the primary constraints multiplied by the Lagrange multiplier 
fields. We assume the standard Poisson brackets for the dynamical variables 
in the  computation for  obtaining the Hamilton's equations of motion. 
We are  led to the following two  
secondary constraints 
\begin{eqnarray}
\Phi\equiv \partial_{-}\pi^{-}+\partial_{\perp}\pi^{\perp}-2\partial_{-}B 
& \approx & 0,  \nonumber \\
\Psi\equiv \pi^{-}+ 2 \partial_{-}A_{+}+\partial_{\perp}A^{\perp}
-(1-\xi)B & \approx & 0. 
\end{eqnarray}
The Hamiltonian is next enlarged by  including  these additional 
constraints as well.   The procedure is repeated.  
No more constraints are seen to arise, and we are left  only  
with the equations 
which would determine the Lagrange multiplier fields. 
According to the 
Dirac procedure \cite {dir1},  
we now go over from the standard Poisson brackets to the  modified Poisson 
 brackets, called frequently Dirac brackets,  such that inside 
 them we are  able to substitute 
the above constraints as  {\it strong} equality 
relations (e.g., by $=$ in place of $\approx$). 
The equal-$\tau$ Dirac bracket $\{f(x),g(y)\}_{D}$ which carries this property 
is constructed straightforwardly. 
 Hamilton's equations now employ the Dirac brackets and    
the   phase space   
constraints $\pi^{+}= 0$, 
$\pi_{B}= 0$, $\eta= 0$, $
\Phi=0$, and $\Psi= 0$ then  effectively reduce the (extended)
Hamiltonian. In the covariant {\it Feynman gauge} with $\xi=1$ 
the free Hamiltonian takes the simple form 
\begin{equation}
{H_{0}}^{LF} = 
-{1\over2}\int {d^{2}x^{\perp}}dx^{-}\; g^{\mu\nu} 
A_{\mu}\,\partial^{\perp}\partial_{\perp}\,A_{\nu}.  
\end{equation}
 The theory is canonically quantized 
through the correspondence $i\{f(x),g(y)\}_{D} \to 
\left[f(x),g(y)\right]$,  the  commutator among the 
corresponding operators. 
 The equal-$\tau$ commutators of the gauge field are found to be 
$\left[A_{\mu}(x),A_{\nu}(y)\right]_{x^{+}=y^{+}=\tau}
=-ig_{\mu\nu} K(x,y)$ 
 where $K(x,y)=-(1/4)\epsilon(x^{-}-y^{-})\delta(x^{\perp}-y^{\perp})$ is 
 nonlocal in the longitudinal coordinate.    
The transverse components of the gauge field have the physical LF 
commutators $ \left[A_{\perp}(x),A_{{\perp}'}(y)\right]_{\tau}
= i \delta_{\perp,\perp'}\,K(x,y)\;$,
while for the $\pm $ components we have only the  mixed 
 commutator nonvanishing $\left[A_{+}(x),A_{-}(y)\right]_{\tau}
 =- i K(x,y)$,  it has a  negative  sign 
 which   indicates  the presence of unphysical 
degrees of freedom in Feynman  gauge.   For        
  $\xi\neq 1$ 
the commutator, for example, 
of $A_{\pm}$ with $A_{\perp}$ is  found  to be nonvanishing.

The Heisenberg
equations of motion lead to ${\Box A_{\mu}}=0$  
for all the components, and consequently  the Fourier transform  
of the free gauge field
over the complete set of plane wave solutions   takes 
 the following form on the LF 
\begin{equation}
A^{\mu}(x)={1\over {\sqrt {(2\pi)^{3}}}}
\int d^{2}k^{\perp}dk^{+}\,
{\theta(k^{+})\over {\sqrt {2k^{+}}}}\,  
e^{\mu (\lambda)}(k)\left[a_{(\lambda)}(k^{+},k^{\perp})
 e^{-ik.x}
+a^{\dag}_{(\lambda)}(k^{+},k^{\perp})
 e^{ik.x} \right ] 
\end{equation}
where  $e^{\mu (\lambda)}(k)$, $\lambda=-,+,1,2\,$ label  
the  set of four linearly independent 
polarization four-vectors. In the {\it front form} theory 
 the two transverse (physical) polarization vector
are space-like while the others are null four-vectors.  
For a fixed  $k^{\mu}=(k^{0},{\vec k})$, where $k^{0}=|{\vec k}|$, 
we may construct them as follows:  
$\,e^{(+)}=(1,{\vec k}/k^{0})/{\sqrt 2}\,$, 
$e^{(-)}=(1,-{\vec k}/k^{0})/{\sqrt 2}\,$, 
$e^{(1)}=(0, {\vec \epsilon }(k;1))\,$, and  
$e^{(2)}=(0, {\vec \epsilon }(k;2))$. Here $(0,1,2,3)$ components are 
specified for convenience while  ${\vec \epsilon }(k;1)$, 
${\vec \epsilon }(k;2)$ and ${\vec k}/|{\vec k}|$ constitute the 
usual orthonormal set of 3-vectors. 
The polarization vectors are orthonormal: 
$g_{\mu\nu}e^{(\lambda){\mu}}(k)e^{(\sigma){\nu}}(k)= 
g^{\lambda\sigma}$ and  satisfy the completeness relation: 
$g_{\lambda\sigma}{e^{(\lambda)}}_{\mu}(k){e^{(\sigma)}}_{\nu}(k)
=  g_{\mu\nu}$. The field commutation relations for the gauge field found above   
are  shown to be satisfied  if we assume, parallel to the 
discussion in the fermionic case, the canonical commutation relations:      
$\,\left[a_{(\lambda)}(k^{+},k^{\perp}),{a^{\dag}}_{(\sigma)}(k'^{+},k'^{\perp})
\right]$ $=-g_{\lambda\sigma}$
$\delta(k^{+}-k'^{+})$ $\delta^{2}(k^{\perp}-k'^{\perp})$. 
We note that   the operators $a_{(0)}=(a_{(+)}+a_{(-)})/{\sqrt 2}$ and 
$a_{(3)}=(a_{(+)}-a_{(-)})/{\sqrt 2}$ obey 
the usual canonical  commutation relations   except that  in the 
case of $a_{(0)}$  a negative 
sign is obtained. The discussion of the 
  Gupta-Bleuler consistency condition  then becomes parallel 
to that in the 
 conventional equal-time treatment \cite{bjk} of the theory.  
 The Feynman gauge free gauge field propagator  on the LF 
 can be derived straightforwardly  using   (20)  
\begin{equation}
 <0| T({A^{a}}_{\mu}(x){A^{b}}_{\nu}(0))|0>= {{i\delta^{ab}}
 \over {(2\pi)^{4}}} 
 \int d^{4}k \;e^{-ik.x}\; 
 {-g_{\mu\nu}\over {k^{2}+i\epsilon}}
 \end{equation}
 where we have restored the gauge  index $a$ and used 
 $\,\left[a_{(\lambda)a}(k^{+},k^{\perp}),{a^{\dag}}
_{(\sigma)b}(k'^{+},k'^{\perp})
\right]$ $=-g_{\lambda\sigma} {\delta}_{ab}
\delta(k^{+}-k'^{+})\delta^{2}(k^{\perp}-k'^{\perp})$.
The  free ghost-antighost  propagators for the anticommuting ghost fields 
may likewise   be derived;  they 
 are also causal and given by  
$(i/(2\pi)^{4}) \delta_{ab}/(k^{2}+i\epsilon)$.)

\section{Illustrations}

The  Hamiltonian density  in Feynman gauge is  
\begin{eqnarray}
{\cal H}^{LF}
&=& {\cal H}_{0}+{\cal H}_{int }\nonumber \\
&=&  -{1\over2} g^{\mu\nu} 
{A^{a}}_{\mu}\,\partial^{\perp}\partial_{\perp}\,{A^{a}}_{\nu}
  -g { \sqrt {2}}{\bar\psi_{+}}^{i}
\gamma^{0}{A_{+}}^{ij}{\psi_{+}}^{j} \nonumber \\
&-& {\bar\psi_{+}}^{i}\left[\delta^{ij}
(i\gamma^{\perp}\partial_{\perp}-m)+g\gamma^{\perp}{A^{ij}}_{\perp}
\right] {\psi_{-}}^{j} 
 +{g\over2}f^{abc}(\partial_{\mu}{A^{a}}_{\nu}-
\partial_{\nu}{A^{a}}_{\mu}) A^{b\mu} A^{c\nu} \nonumber \\
&+&{{g^2}\over 4} 
f^{abe}f^{cde} {A^{a}}_{\mu} {A^{b}}_{\nu} A^{c\mu} A^{d\nu} 
+\partial^{\mu}({\bar\chi}^{a})\partial_{\mu}\chi^{a}
 +g f^{abc} (\partial^{\mu}{\bar\chi^{a}})\chi^{b}{A^{c}}_{\mu}
\end{eqnarray}
where ${\psi_{-}}^{j}$ is given in (4),  we have 
set ${\sqrt2}\chi=({\chi_{1}}+i {\chi_{2}}),\,   
{\sqrt 2}\bar\chi=({\chi_{1}}-i {\chi_{2}})$, and in (22) the cubic and higher
order  terms belong to  ${\cal H}_{int}$ which is also understood to be normal
ordered. It is worth remarking that despite the presence of the 
longitudinal operators  $a_{\pm}$ and ${a^{\dag}}_{\pm}$ 
in the fields $A_{\mu}$, 
there are no non-zero matrix elements involving these quanta
as external lines in view of the commmutation relations of these operators 
as discussed in the previous section.

The  perturbation theory expansion in the interaction representation 
where we time order with respect to the LF time $\tau$ is 
built following the Dyson-Wick \cite{bjk} procedure. We will illustrate 
it in our context through   some explicit  computations,  
for simplicity, in QED where $U(x|A_{-})=exp \{-ie\int_{}^{x^{-}} du^{-}
A_{-}(\tau,u^{-},x^{\perp})\}$ and $D_{\mu}=(\partial_{\mu}-ieA_{\mu})$. 
We observe from (4) and (22) that a {\it seagull} term of the order $e^{2}$ 
is present in the interaction Hamiltonian at the tree level as 
in  case of the scalar field QED. 

Consider  {\it Electron-Muon scattering}.   
The  contribution  coming  from the longitudinal components of the 
gauge field, viz, $A_{+}$ and 
$ A_{-}$, which  carry a nonvanishing mixed  propagator, 
is contained in the following normal-ordered product of the Wick 
decomposition of the second order term in the  perturbation 
theory expansion of the S-matrix \begin{eqnarray}
e^{2} &\int & d^{4}x_{1}d^{4}x_{2}
 :{\psi^{\dag}}_{e +}(x_{1})
 {\psi}_{e +}(x_{1}) {\dot A}_{+}(x_{1})
\;{\psi^{\dag}}_{\mu +}(x_{2}) 
(m+i\not\partial^{T})      \nonumber \\
& &{1\over2}\times\,\int
d{y_{2}}^{-}\epsilon({x_{2}}^{-}-{y_{2}}^{-})
 \{\int_{{y_{2}}^{-}}^{{x_{2}}^{-}}d{u_{2}}^{-} \dot{A}_{-}(u_{2})\}
(m-i\not\partial^{T}) {\psi}_{\mu +}(y_{2}):
\end {eqnarray}
where  overdots indicate \cite{bjk} the  pair of fields contracted, 
$x_{2}\equiv{(\tau_{2}, x_{2}^{-}, x_{2}^{\perp})}$, 
$u_{2}\equiv{(\tau_{2}, u_{2}^{-}, x_{2}^{\perp})}$, 
$y_{2}\equiv{(\tau_{2}, y_{2}^{-}, x_{2}^{\perp})}$,
$x_{1}\equiv{(\tau_{1}, x_{1}^{-}, x_{1}^{\perp})}$, and $\not{\hbox{\kern-4pt $\partial$}}^{T}
\equiv \gamma^{\perp}\partial_{\perp}$.  Here we write $\int \epsilon(x-y) dy/2$ 
in place of ${\partial_{x}}^{-1}$ and suppress the 
symmetry factors  for convenience.  
 Making use of the Fourier 
transforms  of the fields and the gauge field propagator  
 the matrix element is written down by simple inspection 
\begin{equation}
e^{2} \left[{u^{\dag}}_{e +}(p'_{1}){u_{e +}(p_{1})}\;
{u^{\dag}}_{\mu +}(p'_{2}){{(m+{{\not p'}_{2}}^{T})}
\over {-i{{p'}^{+}}_{2}}}{{(m-{{\not p}_{2}}^{T})}
\over {-i{{p}^{+}}_{2}}}u_{\mu +}(p_{2})
\right] \;\;{{-g_{+-}}\over {q^{2}+i\epsilon}}
\end{equation}
where $q^{\mu}=({{{p'}}^{\mu}}_{2}-{{{p}}^{\mu}}_{1})$.  
It may be readily rewritten in view of (10) and the simplifications 
following from the mass shell conditions 
on the external lines  to give  
\begin{equation}
e^{2}\left[{\bar u}_{e }(p'_{1})\gamma^{+}{u_{e }(p_{1})}\;
{\bar u}_{\mu }(p'_{2})\gamma^{-} 
{u_{\mu }(p_{2})}\right]\;\; {{-g_{+-}}\over {q^{2}+i\epsilon}} 
\end{equation}
The  graph with the $A_{+}$ and $A_{-}$ interchanged gives rise to 
a similar expression with  $g_{+-}\to g_{-+}$ while $\gamma ^{\pm}\to 
\gamma ^{\mp}$. There are four contributions arising from the virtual 
propagation  of the transverse components  of the gauge field. 
 They add up to 
\begin{equation}
e^{2}\left[{\bar u}_{e }(p'_{1})\gamma^{\perp}{u_{e }(p_{1})}\;
{\bar u}_{\mu }(p'_{2})\gamma^{\perp'} 
{u_{\mu }(p_{2})}\right]\;\; {{-g_{\perp \perp'}}\over
{q^{2}+i\epsilon}}
\end{equation}
resulting in  
the  following complete matrix element to the second order  
\begin{equation}
e^{2}\left[{\bar u}_{e }(p'_{1})\gamma^{\mu}{u_{e }(p_{1})}\;
{\bar u}_{\mu }(p'_{2})\gamma^{\nu} 
{u_{\mu }(p_{2})}\right]\;\; {{-g_{\mu \nu}}\over {q^{2}+i\epsilon}}.
\end{equation}
which agrees with the one  obtained in the conventional equal-time formulation.

Consider next the  computation of  {\it  Electron Self-Energy}.  
The contribution  from the longitudinal components arises from 
\begin{eqnarray}
e^{2}{\int }d^{4}x_{1} d^{4}x_{2}&&  
: {\psi_{+}}^{\dag}(x_{2})(m+i\not{\partial_{2}}^{T})\int_{-\infty}^{\infty}
{1\over2}d{y_{2}}^{-}\epsilon({x_{2}}^{-}-{y_{2}}^{-}) \nonumber \\
&& \{\int_{{y_{2}}^{-}}^{{x_{2}}^{-}}d{u_{2}}^{-} \dot{A}_{-}(u_{2})\}
(m-i\not{\partial_{2}}^{T}) {\ddot\psi}_{+}(y_{2}){{\ddot\psi}_{+}}^{\dag}(x_{1})
\psi_{+}(x_{1})\dot {A}_{+}(x_{1}):
\end {eqnarray}
leading to 
\begin{equation}
e^{2}\int d^{4}q \;{{{\bar u}^{(r)}(p)[\gamma^{-}(m+\not q^{T})
\gamma^{+}] u^{(s)}(p)}\over
 {[(p-q)^{2}+i\epsilon ]}(q^{2}-m^{2}+i\epsilon)}\; (-g_{-+})
\end{equation}
The  graph with the $A_{+}$ and $A_{-}$ interchanged gives rise to 
a similar expression with  $g_{+-}\to g_{-+}$ while $\gamma ^{\pm}\to 
\gamma ^{\mp}$. 
The matrix elements following from  the  four 
graphs corresponding to the exchange of the 
 ( photon ) fields $A_{1}$ and $A_{2}$  is also
written down by simple inspection.  As in the earlier case  
the expressions get  simplified in virtue of (10) and 
acquire the covariant form encountered in the conventional covariant perturbation 
theory. The   complete matrix element 
  is found to be 
\begin{equation}
e^{2}\int d^{4}q \;{{{\bar u}^{(r)}(p)[\gamma^{\mu}(m+\not{\tilde q})
\gamma^{\nu}]
 u^{(s)}(p)}\over
 {[(p-q)^{2}+i\epsilon ](q^{2}-m^{2}+i\epsilon)}} \;(-g_{\mu\nu})
\end{equation}
where ${{\tilde q}^{\mu}}\equiv( (m^{2}+q^{\perp}q^{\perp})/{(2q^{+})}, 
q^{+}, q^{\perp})$
  and the integration measure is $d^{4}q= d^{2}q^{\perp}dq^{+}dq^{-}$. 
  Considering that the integrand has  the pole at $q^{2}-m^{2}\approx 0$ we may 
regard the expression obtained on the LF to be effectively identical 
to the one obtained in the  conventional covariant theory  framework. The
contribution coming from the seagull term at the tree level vanishes if 
the dimensional regularization is used. 

The calculation of the tree graphs for {\it Compton scattering}, 
$\,\gamma+e\to \gamma+e$, is tedious but straightforward. 
The  results on the LF are shown to be in agreement  
   with the conventional covariant theory one. We remark that 
  on the LF the  tree level {\it seagull} 
term dominates  the (classical)  Thomson formula 
 for the scattering at the vanishingly small  photon energies similar to 
 the case of the QED with the scalar fields.  
  In the covariant gauges on the LF  
 all the relevant field propagators in momentum space  are causal. 
 Employing (10) and (12) it is  rather straightforward 
 to rewrite the  final result in manifestly covariant form. 
   The computation to  higher orders in  the Dyson-Wick perturbation 
expansion can be carried out straightforwardly; the nonlocality 
of the interaction, arising  from (4),  
does require some extra effort to handle, but it 
becomes  easier 
to control the rotational   invariance and compute the loop integrals 
in the traditional fashion.

\section{Conclusions}

 The LF   Dyson-Wick perturbation theory expansion based on the 
LF-time-ordering has a number of advantages for 
computing  the 
high order corrections in the 
{\it front form} QCD theory.  The covariant Feynman (or Landau) gauge  
may be adopted with  advantage on the LF where   all the relevant 
 propagators  take  (except for the numerical projector in the 
 dynamical fermion propagator) the covariant  causal forms 
 permitting us to employ  the usual power counting rules. The 
 Hamiltonian version\cite{ref:cc} can be implemented  in DLCQ \cite{Pauli:1985ps}. In the approximation of no 
 retardation, the Coulomb interaction is recovered. 
  The propagator of  the independent
 component of the fermion field on the LF has been shown  
  to be simpler than in the conventional theory.   
     The   momentum space integrals in the {\it front form} theory 
 may  be  converted \cite{pre2}  to the Euclidean space integrals 
 which  then permit us to 
 employ, for example,  dimensional regularization. 
     The  illustrations given  here  demonstrate   
     the  agreement of the LF quantized 
 theory results with  the conventional covariant theory ones.   
 The expression (10) for  the LF spinor proved to be quite  useful here. 
 The fact that  in the {\it front form} theory the classical 
 Thomson  scattering limit is obtained from a seagull term at the tree level is 
 significant since,  it seems  
  difficult to build on the LF a systematic procedure to obtain 
  semiclassical approximation.
   
 It is worth  remarking also 
 that we have made an {\sl ad hoc} choice of  only 
one (of the family) of  the  characteristic LF hyperplanes,  
 $x^{+}=const.$, in  order to quantize the theory. 
 The conclusions here   confirm  the conjecture \cite{pre1} 
 made earlier on the irrelevance  
 in the quantized theory of the fact that the hyperplanes $x^{\pm}=0$
 constitute  characteristic surfaces of hyperbolic partial differential
 equation. 
   The discussion given in this paper  
 may   clearly be repeated  
for the case of the  non-covariant gauges such as $\partial_{-}A_{-}=0$.  
    It will    appreciably simplify the expression 
of $\psi_{-}$ given in  (4), but now in view of  the 
form of the gauge field propagator, we 
will require   the well known  prescriptions \cite{man} 
in order to recover on the LF the conventional covariant theory results.

\section*{Acknowledgments}

We  acknowledge with thanks the helpful comments  
from Richard Blankenbecler, Sidney  Drell, Michael Peskin, and 
Gerald  Miller during the 
progress of the work. The hospitality offered to PPS at the SLAC and 
a financial grant from the Proci\^encia program of  the UERJ, Rio de Janeiro, 
Brazil,  are gratefully acknowledged.

\end{document}